\documentstyle[11pt,epsfig]{article}
\title{\bf Virial mass in warped DGP-inspired ${\cal L}(R)$ gravity}
\author{{Malihe Heydari-Fard \thanks{E-mail:
heydarifard@qom.ac.ir} \thanks{E-mail: m.heydarifard@sbu.ac.ir}
and Mohaddese Heydari-Fard}\thanks{E-mail: mheydarifard@qom.ac.ir}
\\{\small {{\emph{Department of Physics, The University of Qom, P.
O. Box 37155-1814, Qom, Iran}}}} }
\begin{document}
\maketitle 
\begin{abstract}
A version of the virial theorem is derived in a brane-world
scenario in the framework of a warped DGP model where the action
on the brane is an arbitrary function of the Ricci scalar, ${\cal
L}(R)$. The extra terms in the modified Einstein equations
generate an equivalent mass term (geometrical mass), which give an
effective contribution to the gravitational energy and offer
viable explanation to account for the virial mass discrepancy in
clusters of galaxies. We also obtain the radial velocity
dispersion of galaxy clusters and show that it is compatible with
the radial velocity dispersion profile of such clusters. Finally,
we compare the result of the model with ${\cal L}(R)$ gravity
theories.
\vspace{5mm}
\vspace{0.5mm}
\end{abstract}

\section{Introduction}

The intriguing possibility that our universe is only part of a
higher dimensional space-time has raised a lot of interest in the
physics community \cite{Rubakov}, motivated by the developments in
superstring and M-theory. According to the brane-world scenario,
all matter and gauge interactions reside on the brane, while
gravity can propagate in the whole five-dimensional space– time
(figure 1) \cite{Witten}. Higher dimensional models have a long
history, but it revived by works of Randall and Sundrum (RS) in
1999 \cite{RS}. They have introduced two models in order to solve
the hierarchy problem in particle physics, however after a while
these two models, because of their interesting properties, could
attract a salient attention in cosmology. In their first model
they suppose two brane which our brane has a negative tension
(Shiromizu et al. \cite{SMS} have shown that this model is
unphysical). In their second model they suppose a brane with
infinite extra dimension. In that model our universe has a
positive tension. The effective four-dimensional gravity in the
brane is modified by extra dimension. The cosmological evolution
of such a brane universe has been extensively investigated and
effects such as a quadratic density term in the Friedmann
equations have been found \cite{review1, review2, review3}. An
alternative scenario was subsequently proposed by Dvali, Gabadadze
and Porrati (DGP) \cite{Dvali}. The DGP proposal rests on the key
assumption of the presence of a four-dimensional Ricci scalar in
the brane action. There are two main reasons that make this model
phenomenologically appealing. First, it predicts that
four-dimensional Newtonian gravity on a brane-world is regained at
distances shorter than a given crossover scale $r_c$ (high energy
limit), whereas five-dimensional effects become manifest above
that scale (low energy limit) \cite{Gabadadze}. Second, the model
can explain late-time acceleration without having to invoke a
cosmological constant or quintessential matter \cite{Deffayet}. An
extension of the DGP brane-world scenario have been constructed by
Maeda, Mizuno and Torii which is the combination of the RS II
model and DGP model \cite{maeda}. In this combination, an induced
curvature term appears on the brane in the RS II model which has
been called the warped DGP brane-world in literature \cite{Cai}.
In this paper, we consider the effective gravitational field
equations within the context of the warped DGP brane-world model
where the action on the brane is an arbitrary function of the
Ricci scalar, ${\cal L}(R)$, (which is called the modified DGP
brane-world model in this work) and obtain the spherically
symmetric equations in this scenario. For a recent and
comprehensive review of the phenomenology of DGP cosmology, the
reader is referred to \cite{Lue}.

\begin{figure}
\centerline{\begin{tabular}{ccc}
\epsfig{figure=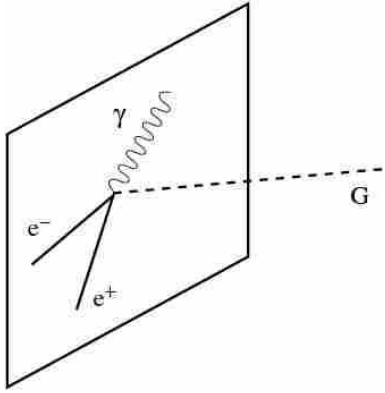,width=5cm}
\end{tabular} } \caption{\footnotesize The confinement of matter to the brane, while gravity propagates in the
bulk.}\label{fig1}
\end{figure}

In the past two decades, various attempts have been made to
understand the nature of the so called dark energy and dark matter
which would be required to explain many observational data. The
question of dark matter is presently one of the most intriguing
aspects in astrophysics and cosmology. There are some compelling
observational evidence for the existence of dark matter for which,
the galaxy rotation curves and mass discrepancy in cluster of
galaxies are two prominent examples \cite{book}. Galaxy clusters
are the largest virialized structures in the universe, and their
mass content is supposed to be representative of the universe as a
whole. The total mass of galaxy clusters can be determined in a
variety of ways. The application of the virial theorem to
positions and velocities of cluster member galaxies is the oldest
method of cluster mass determination \cite{1}. More recent methods
are based on the dynamical analysis of hot x-ray emitting gas
\cite{2} and on the gravitational lensing of background galaxies
\cite{3}. The mass determined from such dynamical means is always
found to be in excess of that which can be attributed to the
visible matter. This is known as the \textbf{\emph{\textbf{missing
mass problem}}}. The existence of dark matter was not firmly
established until the measurement of the rotational velocity of
stars and gases orbiting at a distance $r$ from the galactic
center could be done with reasonable accuracy. Observations show
that the rotational velocity increases near the center of a galaxy
and approaches a nearly constant value with increasing distance
from the center. The discrepancy between the observed rotational
velocity curves and the theoretical prediction from Newtonian
mechanics is known as the \textbf{\emph{\textbf{galactic rotation
curves problem}}} (figure 2). This discrepancy is explained by
postulating that every galaxy and cluster of galaxies is embedded
in a halo made up of some dark matter. To deal with the question
of dark matter, a great number of efforts has been concentrated on
various modifications to the Newtonian gravity or general
relativity \cite{Milgrom, Bekenstein, Mannheim}. There are also
geometrical approaches to address this problem, namely to use
modified Einstein field equations, as is done in brane-world
models \cite{Harko, Shahidi, Heydari, Viznyuk} or in modified
gravity theories \cite{Boehmer, Sefedgar}.

\begin{figure}
\centerline{\begin{tabular}{ccc} \epsfig{figure=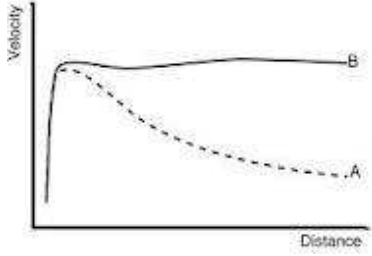,width=5cm}
\end{tabular} } \caption{\footnotesize Observed
rotation velocity curve (solid line B) and the prediction from
Newtonian theory (dotted line A).}\label{fig1}
\end{figure}

In what follows, we first give a brief review of warped DGP
brane-world inspired by the ${\cal L}(R)$ term \cite{Saavedra}.
Next, we introduce the relativistic collision-less Boltzmann
equation from which we deduce the generalized virial theorem with
the aid of the modified Einstein field equations. The geometric
mass and density of a cluster is then identified in terms of the
observable quantities. Then we derive the radial velocity
dispersion relation in galaxies. Finally, in the last section, we
discuss and conclude on our results.

\section{Einstein equations in warped DGP brane-${\cal L}(R)$ model}
We are going to obtain the relativistic virial theorem for galaxy
clusters in warped DGP brane-${\cal L}(R)$ model. For this end we
need, as a first step, to obtain the gravitational field
equations. Consider a 3-brane $\Sigma$ embedded in a
five-dimensional bulk ${\cal M}$. The effective action of a warped
DGP-inspired ${\cal L}(R)$ scenario is given by
\begin{eqnarray}
{\cal S} = {\cal S}_{bulk}+{\cal S}_{brane},\label{1}
\end{eqnarray}
where
\begin{eqnarray}
{\cal S}_{bulk} = \int_{{\cal M}} d^{5}X\sqrt{-{\cal
G}}\left[\frac{1}{2\kappa_5^2}{\cal R}+{\cal
L}_{m}^{(5)}\right],\label{1a}
\end{eqnarray}
\begin{eqnarray}
{\cal S}_{brane} = \int_{\Sigma}d^{4}x \sqrt{-g}{\cal
L}_{brane}(g_{\alpha\beta},\psi)+\int_{\Sigma}d^{4}x
\sqrt{-g}\frac{K^{\pm}}{\kappa_5^2}.\label{1b}
\end{eqnarray}
Here, $X^A$, ($A=0,1,2,3,4$) are five-dimensional coordinates,
while $x^{\mu}$, ($\mu=0,1,2,3$) are the induced four-dimensional
coordinates on the brane. $\kappa_5^2=8\pi G_5$ is the
five-dimensional gravitational constant, ${\cal R}$ and ${\cal
L}_{m}^{(5)}$ are the five-dimensional scalar curvature and the
matter Lagrangian in the bulk, respectively. $K =
\nabla_{\mu}n^\mu$ is the trace of extrinsic curvature on either
side of the brane where $n_{\mu}$ is the unit vecror perpendicular
to the brane hypersurface $\Sigma$ \cite{G}. The last term in
equation (\ref{1b}) is known as the York-Gibbons-Hawking term
which provides a through framework for imposing suitable boundary
conditions on the field equations. ${\cal
L}_{brane}(g_{\alpha\beta},\psi)$ is the effective
four-dimensional Lagrangian, which is given by a generic
functional of the brane metric $g_{\alpha\beta}$ and matter
fields.

The five-dimensional Einstein field equations are given by
\begin{eqnarray}
{\cal R}_{AB}-\frac{1}{2}{\cal R}{\cal G}_{AB} =
{\kappa_5^2}\left[T^{(5)}_{AB}+\delta(Y)\tau_{AB}\right],\label{2}
\end{eqnarray}
where
\begin{eqnarray}
T^{(5)}_{AB}\equiv-2\frac{\delta {\cal L}^{(5)}_{m}}{\delta{\cal
G}^{AB}}+{\cal G}_{AB}{\cal L}^{(5)}_{m},\label{3}
\end{eqnarray}
and
\begin{eqnarray}
\tau_{\mu\nu}\equiv-2\frac{\delta {\cal L}_{brane}}{\delta
g^{\mu\nu}}+g_{\mu\nu}{\cal L}_{brane}.\label{4}
\end{eqnarray}
We study the case where the induced gravity scenario arises from
higher-order corrections on the scalar curvature over the brane.
The interaction between the bulk gravity and local matter induces
gravity on the brane through its quantum effects. If we take into
account quantum effects of the matter fields confined to the
brane, the gravitational action on the brane is modified as
\begin{eqnarray}
{\cal L}_{brane}(g_{\alpha\beta},\psi) = \frac{\mu^2}{2}{\cal
L}(R)-\lambda_b+{\cal L}_{m},\label{5}
\end{eqnarray}
where $\mu$ is a mass scale which may correspond to the
four-dimensional Planck mass, $\lambda_b$ is the tension of the
brane and ${\cal L}_m$ presents the Lagrangian of the matter
fields on the brane. We note that for ${\cal L}(R)=R$, action
(\ref{1}) gives the DGP model if $\lambda_b=\Lambda^{(5)}=0$ and
gives the RSII model if $\mu=0$.

We obtain the gravitational field equations on the brane-world as
\cite{SMS}
\begin{eqnarray}
G_{\mu\nu} = \frac{2\kappa_5^2}{3}\left[T_{AB}^{(5)}g^A_{\mu}
g^B_{\nu}+g_{\mu\nu}\left(T_{AB}^{(5)}n^An^B-\frac{1}{4}T^{(5)}\right)\right]+\kappa_{5}^4\pi_{\mu\nu}-{\cal
E}_{\mu\nu},\label{6}
\end{eqnarray}
where the quadratic correction has the form
\begin{eqnarray}
\pi_{\mu\nu} =
-\frac{1}{4}\tau_{\mu\alpha}\tau_{\nu}^{\alpha}+\frac{1}{12}\tau\tau_{\mu\nu}+
\frac{1}{8}g_{\mu\nu}\tau^{\alpha\beta}\tau_{\alpha\beta}-\frac{1}{24}g_{\mu\nu}\tau^2,\label{8}
\end{eqnarray}
and
\begin{eqnarray}
{\cal E}_{\mu\nu} = C^{(5)}_{ABCD}n^An^Bg^C_{\,\,\,\mu}
g^D_{\,\,\,\nu},\label{E}
\end{eqnarray}
is the projection of the bulk Weyl tensor to the surface
orthogonal to $n^A$. By the variation over the metric, we can also
obtain the surface equation as
\begin{eqnarray}
\kappa_5^2\tau_{\mu\nu} = K_{\mu\nu}-Kg_{\mu\nu},\label{K}
\end{eqnarray}
where $K_{\mu\nu}$ is the extrinsic curvature and $\tau_{\mu\nu}$
is the total energy-momentum tensor of the brane. Equation
(\ref{K}) is nothing but the Israel junction condition
\cite{Israel}. By taking the divergence of the Israel equation, we
have
\begin{eqnarray}
\kappa_5^2\nabla_{\mu}\tau^{\mu\nu} =
\nabla_{\mu}K^{\mu\nu}-g^{\mu\nu}\nabla_{\mu}K,\label{KK}
\end{eqnarray}
where $\nabla_{\mu}$ is the covariant derivative with respect to
$g_{\mu\nu}$. The right hand side can be evaluated using Codacci's
equation and the field equation of the bulk, giving
\begin{eqnarray}
\nabla_{\nu}\tau_{\mu}^{\nu} =
-2T_{AB}^{(5)}n^{A}g^{B}_{\mu}.\label{7}
\end{eqnarray}
Also, we note that the junction conditions are ambiguous and do
not define the four-dimensional geometry in the unique way. If one
changes the surface action (freedom in the choice of boundary
terms), the junction condition (\ref{K}) is also changed. This
shows that brane-world is lacking the physical principle to
predict in unique way the surface term and, hence, the emerging
brane cosmology \cite{junction1}. One may prescribe the AdS/CFT
correspondence to define the boundary action \cite{junction2}
where curvature surface terms correspond to usual counterterms in
dual QFT.

In order to find the basic field equations on the brane with
induced gravity described by ${\cal L}(R)$ term, we have to obtain
the energy-momentum tensor of the brane $\tau_{\mu\nu}$ by
definition (\ref{4}) from Lagrangian (\ref{5}), yielding
\begin{eqnarray}
\tau_{\mu\nu} =
-\Lambda(R)g_{\mu\nu}+T_{\mu\nu}-\Sigma(R)G_{\mu\nu}+\Theta_{\mu\nu},\label{10}
\end{eqnarray}
where the functions $\Lambda(R)$, $\Sigma(R)$ and
$\Theta_{\mu\nu}$ are defined as
\begin{equation}
\Lambda(R) = \frac{\mu^2}{2}\left[R\frac{d{\cal L}(R)}{dR}-{\cal
L}(R)+2\frac{\lambda_b}{\mu^2}\right],\label{11}
\end{equation}
\begin{equation}
\Sigma(R) = \mu^2\frac{d{\cal L}(R)}{dR},\label{12}
\end{equation}
and
\begin{equation}
\Theta_{\mu\nu} = \mu^2\left[\nabla_\mu
\nabla_\nu{\left(\frac{d{\cal
L}(R)}{dR}\right)}-g_{\mu\nu}\nabla^\beta
\nabla_\beta\left(\frac{d{\cal L}(R)}{dR}\right)\right].\label{13}
\end{equation}
Inserting equations (\ref{10}) into equation (\ref{6}), we find
the effective field equations for the four-dimensional metric
$g_{\mu\nu}$ as \cite{Saavedra}
\begin{eqnarray}
&&\left[1+\frac{1}{6}\kappa_{5}^{4}\Lambda(R)\Sigma(R)\right]G_{\mu\nu}
= \frac{1}{6}\kappa_{5}^{4}\Lambda(R)T_{\mu\nu}-\Lambda_4 (R)
g_{\mu\nu}+
\frac{1}{6}\kappa_{5}^{4}\Lambda(R)\Theta_{\mu\nu}+\kappa_{5}^{4}\pi_{\mu\nu}^{(T)}+\kappa_{5}^4\pi_{\mu\nu}^{(\Theta)}\nonumber\\
&+&\kappa_{5}^{4}\Sigma(R)^2\pi_{\mu\nu}^{(G)}-\kappa_{5}^{4}\Sigma(R){
K}^{(T)}_{\mu\nu\rho\sigma}G^{\rho\sigma}-\kappa_{5}^{4}\left[\Sigma(R)G^{\rho\sigma}-T^{\rho\sigma}\right]{
K}^{(\Theta)}_{\mu\nu\rho\sigma}-{\cal E}_{\mu\nu},\label{15}
\end{eqnarray}
where
\begin{eqnarray}
\pi^{(T)}_{\mu\nu} =
-\frac{1}{4}T_{\mu\alpha}T^{\alpha}_{\nu}+\frac{1}{12}TT_{\mu\nu}
+\frac{1}{8}g_{\mu\nu}T_{\alpha\beta}T^{\alpha\beta}-\frac{1}{24}g_{\mu\nu}T^2,\label{17}
\end{eqnarray}
\begin{eqnarray}
\pi^{(G)}_{\mu\nu} =
-\frac{1}{4}G_{\mu\alpha}G^{\alpha}_{\nu}+\frac{1}{12}GG_{\mu\nu}
+\frac{1}{8}g_{\mu\nu}G_{\alpha\beta}G^{\alpha\beta}-\frac{1}{24}g_{\mu\nu}G^2,\label{18}
\end{eqnarray}
\begin{eqnarray}
\pi^{(\Theta)}_{\mu\nu} =
-\frac{1}{4}\Theta_{\mu\alpha}\Theta^{\alpha}_{\nu}+\frac{1}{12}\Theta\Theta_{\mu\nu}
+\frac{1}{8}g_{\mu\nu}\Theta_{\alpha\beta}\Theta^{\alpha\beta}-\frac{1}{24}g_{\mu\nu}\Theta^2,\label{19}
\end{eqnarray}
and
\begin{eqnarray}
{K}^{(T)}_{\mu\nu\rho\sigma} =
\frac{1}{4}\left(g_{\mu\nu}T_{\rho\sigma}-g_{\mu\rho}T_{\nu\sigma}-g_{\nu\sigma}T_{\mu\rho}\right)+\frac{1}{12}
\left[T_{\mu\nu}g_{\rho\sigma}+T(g_{\mu\rho}g_{\nu\sigma}-g_{\mu\nu}g_{\rho\sigma})\right],\label{20}
\end{eqnarray}
\begin{eqnarray}
{K}^{(\Theta)}_{\mu\nu\rho\sigma} =
\frac{1}{4}\left(g_{\mu\nu}\Theta_{\rho\sigma}-g_{\mu\rho}\Theta_{\nu\sigma}-g_{\nu\sigma}\Theta_{\mu\rho}\right)+\frac{1}{12}
\left[\Theta_{\mu\nu}g_{\rho\sigma}+\Theta(g_{\mu\rho}g_{\nu\sigma}-g_{\mu\nu}g_{\rho\sigma})\right],\label{21}
\end{eqnarray}
with $T$ being the trace of the energy-momentum tensor and
$\Theta$ is
\begin{equation}
\Theta = g^{\mu\nu}\Theta_{\mu\nu} =
-3\mu^2\nabla^{\alpha}\nabla_{\alpha}\left(\frac{d{\cal
L}(R)}{dR}\right),\label{22}
\end{equation}
and the effective cosmological constant on the brane is given by
\begin{equation}
\Lambda_4(R) =
\frac{1}{2}\kappa_{5}^{2}\left[\Lambda^{(5)}+\frac{1}{6}\kappa_{5}^{2}\Lambda(R)^2\right].\label{23}
\end{equation}
We can also recover the standard four-dimensional ${\cal L}(R)$
gravity \cite{L(R)} from equation (\ref{15}) in the limit
$\lambda_b^{-1}\rightarrow0$. A alternative possibility in
recovering the four-dimensional ${\cal L}(R)$ gravity is to take
the limit $\kappa_5\rightarrow0$, while keeping the Newtonian
gravitational constant $\kappa_4^2$ finite \cite{SMS} (see
Appendix A).

Now we are interested in computing the spherically symmetric
solutions on the brane. We consider an isolated and spherically
symmetric cluster being described by a static and spherically
symmetric metric
\begin{eqnarray}
ds^2 =
-e^{\lambda(r)}dt^2+e^{\nu(r)}dr^2+r^2(d\theta^2+sin^2\theta
d\varphi^2),\label{24}
\end{eqnarray}
we also assume that the matter content of the bulk is just a
cosmological constant $\Lambda^{(5)}$ and the matter content of
the 3-brane universe is considered to be a cosmological constant
plus a localized spherically symmetric perfect fluid
$$T_{\mu\nu} =
(p_b+\rho_b)v_\mu v_\nu+p_bg_{\mu\nu}.$$ Therefore, the
gravitational field equations become
\begin{eqnarray}
&&\left[1+\frac{1}{6}\kappa_{5}^4\Lambda(R)\Sigma(R)\right]\frac{e^{-\nu}}{r^2}\left(-1+r\nu^{'}+e^{\nu}\right)
=
\frac{1}{6}\kappa_{5}^{4}\Lambda(R)\rho_b+\Lambda_4(R)-\frac{1}{6}\kappa_{5}^4\Lambda(R)\Theta_0^0+\frac{\kappa_{5}^4}{12}\rho_b^2\nonumber\\
&+&\kappa_{5}^{4}\Sigma(R){\cal
K}_0^{0(T)}-\kappa_{5}^4\Sigma(R)^2\pi_0^{0(G)}
-\kappa_{5}^{4}{\cal K}_0^{0(\Theta)}+\kappa_{5}^{4}\Sigma(R){\cal
K}_0^{0(G)}-\kappa_{5}^4\pi_0^{0(\Theta)}+\frac{6}{\kappa_4^2\lambda_b}U(r)
,\label{a25}
\end{eqnarray}
\begin{eqnarray}
&&\left[1+\frac{1}{6}\kappa_{5}^4\Lambda(R)\Sigma(R)\right]\frac{e^{-\nu}}{r^2}\left(1+r\lambda^{'}-e^{\nu}\right)
=
\frac{1}{6}\kappa_{5}^{4}\Lambda(R)p_b-\Lambda_4(R)+\frac{1}{6}\kappa_{5}^4\Lambda(R)\Theta_1^1+
\frac{\kappa_{5}^4}{12}\left(\rho_b^2+2 \rho_b
p_b\right)\nonumber\\
&-&\kappa_{5}^{4}\Sigma(R){\cal K}_1^{1(T)}
+\kappa_{5}^4\Sigma(R)^2\pi_1^{1(G)}+\kappa_{5}^{4}{\cal
K}_1^{1(\Theta)}-\kappa_{5}^{4}\Sigma(R){\cal
K}_1^{1(G)}+\kappa_{5}^4\pi_1^{1(\Theta)}+\frac{2}{\kappa_4^2\lambda_b}\left[U(r)+2P(r)\right]
,\label{a26}
\end{eqnarray}
\begin{eqnarray}
&&\left[1+\frac{1}{6}\kappa_{5}^4\Lambda(R)\Sigma(R)\right]\frac{e^{-\nu}}{2r}\left(2\lambda^{'}-2\nu^{'}-\lambda^{'}\nu^{'}
r+2\lambda^{''}r+\lambda^{'2}r\right) =
\frac{1}{3}\kappa_{5}^{4}\Lambda(R)p_b-2\Lambda_4(R)+\frac{1}{3}\kappa_{5}^4\Lambda(R)\Theta_2^2\nonumber\\
&+&\frac{\kappa_{5}^4}{6}\left(\rho_b^2 +2\rho_b p_b\right)
-2\kappa_{5}^{4}\Sigma(R){\cal
K}_2^{2(T)}+2\kappa_{5}^4\Sigma(R)^2\pi_2^{2(G)}
+2\kappa_{5}^{4}{\cal
K}_2^{2(\Theta)}-2\kappa_{5}^{4}\Sigma(R){\cal
K}_2^{2(G)}\nonumber\\
&+&2\kappa_{5}^4\pi_2^{2(\Theta)}+\frac{4}{\kappa_4^2\lambda_b}[U(r)-P(r)]
,\label{a27}
\end{eqnarray}
where $U(r)$ and $P(r)$ are the dark radiation and dark pressure.
Also ${\cal K}_{\mu\nu}^{(T)}$, ${\cal K}_{\mu\nu}^{(G)}$ and
${\cal K}_{\mu\nu}^{(\Theta)}$ are defined as
$${K}_{\mu\nu\rho\sigma}^{(T)}G^{\rho\sigma} \equiv {\cal
K}_{\mu\nu}^{(T)},$$
$${K}_{\mu\nu\rho\sigma}^{(\Theta)}G^{\rho\sigma} \equiv {\cal
K}_{\mu\nu}^{(G)},$$
$${K}_{\mu\nu\rho\sigma}^{(\Theta)}T^{\rho\sigma} \equiv {\cal
K}_{\mu\nu}^{(\Theta)}.$$ Now, using these relations and adding
the gravitational field equations (\ref{a25})-(\ref{a27}) we find
\begin{eqnarray}
&&\left(1+\frac{1}{6}\kappa_{5}^4\Lambda(R)\Sigma(R)\right)e^{-\nu}\left(\frac{\lambda^{'}}{r}-\frac{\lambda^{'}\nu^{'}}{4}
+\frac{\lambda^{''}}{2}+ \frac{\lambda^{'2}}{4}\right) =
\frac{\kappa_5^4}{12}\Lambda(R)\left(\rho_b+3p_b\right)-\Lambda_4(R)+\frac{\kappa_5^4}{12}\rho_b(2\rho_b+3p_b)\nonumber\\
&-&\frac{1}{12}\kappa_{5}^4\Lambda(R)[\Theta_0^0-\Theta_1^1-2\Theta_2^2]+\frac{\kappa_5^4\Sigma(R)}{2}[{\cal
K}_0^{0(T)}-{\cal K}_1^{1(T)}-2{\cal
K}_2^{2(T)}]+\frac{\kappa_5^4\Sigma(R)}{2}[{\cal K}_0^{0(G)}-{\cal
K}_1^{1(G)}-2{\cal
K}_2^{2(G)}]\nonumber\\
&-&\frac{\kappa_5^4}{2}[{\cal K}_0^{0(\Theta)}-{\cal
K}_1^{1(\Theta)}-2{\cal
K}_2^{2(\Theta)}]-\frac{\kappa_5^4\Sigma(R)^2}{2}[\pi_0^{0(G)}
-\pi_1^{1(G)}-2\pi_2^{2(G)}]\nonumber\\
&-&\frac{\kappa_5^4}{2}[\pi_0^{0(\Theta)}
-\pi_1^{1(\Theta)}-2\pi_2^{2(\Theta)}]+\frac{6}{\kappa_4^2\lambda_b}U(r),\label{28}
\end{eqnarray}
where
\begin{eqnarray}
{\cal K}_0^{0(T)}-{\cal K}_1^{1(T)}-2{\cal K}_2^{2(T)} &=&
\frac{1}{12}e^{-\nu}[\frac{6\rho_b}{r^2}(1-e^{\nu})+\frac{6p_b}{r^2}(1-e^{\nu})
-\frac{4p_b\nu'}{r}+\rho_b\lambda'\nu'-2\rho_b\lambda''-\rho_b\lambda'^2\nonumber\\
&-&\frac{4\rho_b\lambda'}{r}-\frac{2p_b\nu'}{r}-\frac{6\rho_b\nu'}{r}],\label{29}
\end{eqnarray}
\begin{eqnarray}
{\cal K}_0^{0(G)} -{\cal K}_1^{1(G)}-2{\cal K}_2^{2(G)} &=&
\frac{\mu^2}{24}e^{-\nu}\left[(\frac{6\lambda'}{r^2}
+\frac{2\nu'}{r^2}-\frac{8}{r^3})R'\frac{d^2{\cal
L}}{dR^2}-\frac{4}{r^2}R''\frac{d^2{\cal
L}}{dR^2}-\frac{4}{r^2}R'^2\frac{d^3{\cal L}}{dR^3}\right]
\nonumber\\
&-&\frac{\mu^2}{24}
e^{-2\nu}\{(-\frac{8}{r^3}+\frac{10\nu'}{r^2}-{\nu'\lambda'^2}
-\frac{14\nu'\lambda'}{r}+\frac{8\lambda''}{r}
+\lambda'\nu'^2+\frac{22\lambda'}{r^2}+\frac{4\lambda'^2}{r}\nonumber\\
&-&2\nu'\lambda''-\frac{2\nu'^2}{r})R'\frac{d^2{\cal L}}{dR^2}
+(2\lambda'^2-\frac{4}{r^2}+\frac{8\lambda'}{r}+\frac{4\nu'}{r}-2\lambda'\nu'+4\lambda'')R''\frac{d^2{\cal
L}}{dR^2}\nonumber\\
&+&(\frac{4\nu'}{r}+\frac{8\lambda'}{r}-2\lambda'\nu'+2\lambda'^2-\frac{4}{r^2}+4\lambda'')R'^2\frac{d^3{\cal
L}}{dR^3} \},\label{30}
\end{eqnarray}
\begin{eqnarray}
{\cal K}_0^{0(\Theta)} -{\cal K}_1^{1(\Theta)}-2{\cal
K}_2^{2(\Theta)} &=& -\frac{1}{12}\mu^2
e^{-\nu}[\frac{4}{r}(2\rho_b+3p_b)R'\frac{d^2{\cal
L}}{dR^2}-(2\rho_b+3p_b)\nu'R'\frac{d^2{\cal L}}{dR^2}
-3\rho_b\lambda'R'\frac{d^2{\cal
L}}{dR^2}\nonumber\\
&+&2(2\rho_b+3p_b)R''\frac{d^2{\cal
L}}{dR^2}+2(2\rho_b+3p_b)R'^2\frac{d^3{\cal L}}{dR^3}],\label{31}
\end{eqnarray}
and
\begin{eqnarray}
\pi_0^{0(G)} -\pi_1^{1(G)}-2\pi_2^{2(G)} &=&
-\frac{1}{12}e^{-2\nu}[\frac{2\nu'\lambda''}{r}+\frac{\lambda'^2\nu'}{r}+\frac{5\lambda'\nu'}{r^2}-\frac{\lambda'\nu'^2}{r}+
\frac{2}{r^4}-\frac{4\nu'}{r^3}-\frac{\lambda'^2}{r^2}-\frac{4\lambda'}{r^3}
-\frac{2\lambda''}{r^2}\nonumber\\
&+&\frac{2\nu'^2}{r^2}]
-\frac{1}{12}e^{-\nu}[\frac{4\lambda'}{r^3}+\frac{4\nu'}{r^3}+\frac{\lambda'^2}{r^2}+\frac{2\lambda''}{r^2}
-\frac{4}{r^4}-\frac{\lambda'\nu'}{r^2}+\frac{2e^{\nu}}{r^4}],\label{32}
\end{eqnarray}
\begin{eqnarray}
{\pi}_0^{0(\Theta)}-{\pi}_1^{1(\Theta)}-2{\pi}_2^{2(\Theta)} =
\frac{1}{8}e^{-2\nu}\mu^4R'\lambda'\frac{d^2{\cal
L}}{dR^2}\left[(2R''- \nu'R'+\frac{4R'}{r})\frac{d^2{\cal
L}}{dR^2}+2R'^2\frac{d^3{\cal L}}{dR^3}\right],\label{33}
\end{eqnarray}
\begin{eqnarray}
\Theta_0^0-\Theta_1^1-2\Theta_2^2 =
\mu^2e^{-\nu}\left[(R''-\frac{1}{2}\nu'R'+\frac{3}{2}\lambda'R'+\frac{2}{r}R')\frac{d^2{\cal
L}}{dR^2}+ R'^2\frac{d^3{\cal L}}{dR^3}\right],\label{34}
\end{eqnarray}
where a prime represents differentiation with respect to $r$. In
the next section, we will investigate the influence of the bulk
effects on the dynamics of the galaxies in modified DGP
brane-world model.

\section{The virial theorem}

To derive the virial theorem in the context of the model discussed
above we have to first introduce the following frame of
orthonormal vectors \cite{tetrad, Tetrad, Jackson}
\begin{eqnarray}
e^{(0)}_{\rho} = e^{\lambda/2}\delta^{0}_{\rho},\hspace{.5
cm}e^{(1)}_{\rho} = e^{\nu/2}\delta^{1}_{\rho},\hspace{.5
cm}e^{(2)}_{\rho} = r\delta^{2}_{\rho},\hspace{.5
cm}e^{(3)}_{\rho} = r\sin\theta\delta^{3}_{\rho},\label{b1}
\end{eqnarray}
where $g^{\mu\nu}e_{\mu}^{(a)}e_{\nu}^{(b)}=\eta^{(a)(b)}$. The
four-velocity $v^{\mu}$ of a typical galaxy with
$v^{\mu}v_{\mu}=-1$, in tetrad components is written as
\begin{eqnarray}
v^{(a)} = v^{\mu}e^{(a)}_{\mu},\hspace{.5 cm} a =
0,1,2,3.\label{b2}
\end{eqnarray}
Now, we write down the general relativistic Boltzmann equation
governing the evolution of the distribution function of galaxies
$f_B$, which are treated as identical and collision-less point
particles. This relativistic Boltzmann equation in tetrad
components is given by
\begin{eqnarray}
v^{(a)}e^{\rho}_{(a)}\frac{\partial f_B}{\partial
x^{\rho}}+\gamma^{(a)}_{(b)(c)}v^{(b)}v^{(c)}\frac{\partial
f_B}{\partial v^{(a)}} = 0,\label{b3}
\end{eqnarray}
where $f_B = f_B(x^{\mu},v^{(a)})$ and
$\gamma^{(a)}_{(b)(c)}=e^{(a)}_{\rho;\sigma}e^{\rho}_{(b)}e^{\sigma}_{(c)}$
are the distribution function and the Ricci rotation coefficients,
respectively. Assuming that the distribution function is only a
function of $r$, the relativistic Boltzmann equation becomes
\begin{eqnarray}
v_{r}\frac{\partial f_B}{\partial
r}&-&\left(\frac{v_{t}^{2}}{2}\frac{\partial \lambda}{\partial
r}-\frac{v_{\theta}^{2}+v_{\varphi}^{2}}{r}\right)\frac{\partial
f_B}{\partial v_{r}}-\frac{v_{r}}{r}\left(v_{\theta}\frac{\partial
f_B}{\partial v_{\theta}}+v_{\varphi}\frac{\partial f_B}{\partial
v_{\varphi}}\right)\nonumber\\
&-&\frac{e^{\nu/2}v_{\varphi}}{r}\cot{\theta}\left(v_{\theta}\frac{\partial
f_B}{\partial v_{\varphi}}-v_{\varphi}\frac{\partial f_B}{\partial
v_{\theta}}\right) = 0,\label{b4}
\end{eqnarray}
where
\begin{equation}\label{}
v^{(0)} = v_t,\hspace{.5 cm}v^{(1)} = v_r,\hspace{.5 cm}v^{(2)} =
v_\theta,\hspace{.5 cm}v^{(3)} = v_\varphi.\label{b5}
\end{equation}
Since our metric is spherically symmetric, the coefficient of
cot$\theta$ must be zero. Multiplying equation (\ref{b4}) by $m
v_r dv$ where $dv = \frac{1}{v_t}dv_r dv_\theta dv_\varphi$ is the
invariant volume element in the velocity space and $m$ is the mass
of the galaxy, and integrating over the velocity space and
assuming that the distribution function vanishes rapidly as the
velocities tend to $\pm \infty$, we find
\begin{eqnarray}
r\frac{\partial}{\partial r}\left[\rho\langle { v_r^2 }
\rangle\right]+\frac{1}{2}\rho\left[\langle{v_t^2}\rangle+\langle{v_r^2}\rangle\right]
r\frac{\partial \lambda}{\partial
r}-\rho\left[\langle{v_\theta^2}\rangle+\langle{v_\varphi^2}\rangle-2\langle{v_r^2}\rangle\right]
 = 0,\label{b6}
\end{eqnarray}
where $\rho$ is the mass density and $\langle v_r^2\rangle$
represents the average value of $v_r^2$. Multiplying equation
(\ref{b6}) by $4\pi r^2$ and integrating over the cluster leads to
\begin{eqnarray}
&&\int^{R}_{0}4\pi\rho\left[\langle{v_r^2}\rangle
+\langle{v_\theta^2}\rangle+\langle{v_\varphi^2}\rangle\right]r^2
dr-\frac{1}{2}\int^{R}_{0}4\pi r^3\rho
\left[\langle{v_t^2}\rangle+\langle{v_r^2}\rangle\right]\frac{\partial
\lambda}{\partial r}dr = 0.\label{b7}
\end{eqnarray}
This equation can be written as
\begin{eqnarray}
2K-\frac{1}{2}\int^{R}_{0}4\pi
r^3\rho\left[\langle{v_t^2}\rangle+\langle{v_r^2}\rangle\right]\frac{\partial
\lambda}{\partial r}dr = 0,\label{b8}
\end{eqnarray}
where the total kinetic energy of the galaxies is defined as
\begin{eqnarray}
K = \int^{R}_{0}2\pi\rho\left[\langle{v_r^2}\rangle+
\langle{v_\theta^2}\rangle+\langle{v_\varphi^2}\rangle\right]r^2
dr.\label{b9}
\end{eqnarray}
To obtain the virial theorem in our model we must express the
energy-momentum tensor components in the terms of the distribution
function. This is done according to
\begin{eqnarray}
T_{\mu\nu} = \int f_B m v_{\mu}v_{\nu} dv,\label{b10}
\end{eqnarray}
which leads to
\begin{eqnarray}
\rho_{b} = \rho\langle{v_{t}^{2}}\rangle,\hspace{.5 cm}p_{b} =
\rho\langle{v_{r}^{2}}\rangle = \rho\langle{v_{\theta}^{2}}\rangle
= \rho\langle{v_\varphi^{2}}\rangle.\label{b11}
\end{eqnarray}
Now, using these relations and equation (\ref{28}) we obtain
\begin{eqnarray}
&&\left(1+\frac{1}{6}\kappa_{5}^4\Lambda(R)\Sigma(R)\right)e^{-\nu}\left(\frac{\lambda^{'}}{r}-\frac{\lambda^{'}\nu^{'}}{4}
+\frac{\lambda^{''}}{2}+ \frac{\lambda^{'2}}{4}\right) =
\frac{\kappa_5^4}{12}\Lambda(R)\rho[\langle{v_t^2}\rangle+\langle{v_r^2}\rangle+
\langle{v_\theta^2}\rangle+\langle{v_\varphi^2}\rangle]\nonumber\\
&-&\Lambda_4(R)
+\frac{\kappa_5^4}{12}\rho^2[\langle{v_t^2}\rangle^2+\langle{v_r^2}\rangle^2+
\langle{v_\theta^2}\rangle^2+\langle{v_\varphi^2}\rangle^2]
-\frac{\kappa_{5}^4}{12}\Lambda(R)[\Theta_0^0-\Theta_1^1-2\Theta_2^2]\nonumber\\
&+&\frac{\kappa_5^4\Sigma(R)}{2}[{\cal K}_0^{0(T)}-{\cal
K}_1^{1(T)}-2{\cal K}_2^{2(T)}] -\frac{\kappa_5^4}{2}[{\cal
K}_0^{0(\Theta)}-{\cal K}_1^{1(\Theta)}-2{\cal
K}_2^{2(\Theta)}]\nonumber\\
&+&\frac{\kappa_5^4\Sigma(R)}{2}[{\cal K}_0^{0(G)}-{\cal
K}_1^{1(G)}-2{\cal K}_2^{2(G)}]
-\frac{\kappa_5^4\Sigma(R)^2}{2}[\pi_0^{0(G)}
-\pi_1^{1(G)}-2\pi_2^{2(G)}]\nonumber\\
&-&\frac{\kappa_5^4}{2}[\pi_0^{0(\Theta)}
-\pi_1^{1(\Theta)}-2\pi_2^{2(\Theta)}]+\frac{6}{\kappa_4^2\lambda_b}U(r).\label{b12}
\end{eqnarray}
In order to obtain the virial theorem we have also to use some
approximations. First, consider that the galaxies in the cluster
have velocities much smaller than the velocity of light. In other
words $$\langle{v_r^2}\rangle\approx
\langle{v_\theta^2}\rangle\approx
\langle{v_\varphi^2}\rangle\ll\langle{v_t^2}\rangle\approx1.$$
Second, consider inside the galactic clusters the gravitational
field is weak and we can  use the weak gravitational field
approximation. Then the term proportional to $\lambda'\nu'$ and
${\lambda'}^{2}$ in equation (\ref{b12}) may be ignored. Third,
since we are interested in astrophysical applications at the
extra-galactic scale, we assume that the deviations from standard
general relativity (corresponding to the background value
$\frac{d{\cal L}}{dR}=1$) are small. Therefore we may represent
$\frac{d{\cal L}}{dR}$ as $\frac{d{\cal L}}{dR} = 1+\varepsilon
g'(R)$, where $\varepsilon$ is a small quantity and $g'(R)$
describes the modifications of the geometry due to the presence of
the tensor $\Theta_{\mu\nu}$ \cite{Boehmer}. Thus, assuming that
$e^{-\lambda}\approx 1$ inside the cluster \cite{Harko}, equations
(\ref{30}), (\ref{32}) and (\ref{33}) are zero and equation
(\ref{b12}) reduces to
\begin{eqnarray}
&&\left[1+\frac{1}{6}\kappa_{5}^4\lambda_b\left(1+\frac{1}{2\lambda_b}\mu^2\varepsilon
g'(R)\right)\mu^2\left(1+\varepsilon
g'(R)\right)\right]\left(\frac{\lambda^{'}}{r}
+\frac{\lambda^{''}}{2}\right) =
\frac{\kappa_5^4}{12}\lambda_b\left(1+\frac{1}{2\lambda_b}\mu^2\varepsilon
g'(R)\right)\rho\nonumber\\
&\times&\left[\langle{v_t^2}\rangle+\langle{v_r^2}\rangle+
\langle{v_\theta^2}\rangle+\langle{v_\varphi^2}\rangle\right]
-\Lambda_4(R)+\frac{\kappa_4^2}{2\lambda_b}\rho^2\left[\langle{v_t^2}\rangle^2+\langle{v_r^2}\rangle^2+
\langle{v_\theta^2}\rangle^2+\langle{v_\varphi^2}\rangle^2\right]\nonumber\\
&-&\frac{1}{12}\kappa_{5}^4\mu^2\lambda_b\left(1+\frac{1}{2\lambda_b}\mu^2\varepsilon
g'(R)\right)(R_0''+\frac{2}{r}R_0')\frac{d^2{\cal L}}{dR^2}\nonumber\\
&-&\frac{\kappa_5^4}{6r}\mu^2\left(1+\varepsilon
g'(R)\right)\rho\left[3\nu'\langle{v_t^2}\rangle+2\lambda'\langle{v_t^2}\rangle
+\lambda''
r\langle{v_t^2}\rangle+3\nu'\langle{v_r^2}\rangle\right] \nonumber\\
&+&\frac{1}{12}\kappa_{5}^4\mu^2\rho\left[2\langle{v_t^2}\rangle+\langle{v_r^2}\rangle+
\langle{v_\theta^2}\rangle+\langle{v_\varphi^2}\rangle\right](R_0''+\frac{2}{r}R_0')\frac{d^2{\cal
L}}{dR^2} +\frac{6}{\kappa_4^2\lambda_b}U(r),\label{b13}
\end{eqnarray}
where $R = R_0+{\cal{O}}({{\lambda'}^2},\lambda'\nu')$. On the
other hand, for clusters of galaxies the ratio of the matter
density and of the brane tension is much smaller than 1,
$\rho/\lambda_b << 1$, so that one can neglect the quadratic term
in the matter density in above equation. These conditions
certainly apply to test particles in stable circular motion around
galaxies and to the galactic clusters. Thus, we can rewrite
equation (\ref{b13}) as
\begin{eqnarray}
\left(1+\frac{\lambda_b}{6}\kappa_{5}^4\mu^2\right)\frac{1}{2r^2}\frac{\partial}{\partial
r}\left(r^2\frac{\partial \lambda}{\partial r}\right) =
\frac{\kappa_4^2}{2}\rho(r)-\Lambda_4+\frac{\kappa_4^2}{2}\rho_{curv}(r)+{\kappa_5^4\mu^2}[{\cal
P}(r)+3{\cal U}(r)]+\frac{6}{\kappa_4^2\lambda_b}U(r),\label{b14}
\end{eqnarray}
where
\begin{eqnarray}
{\cal U}(r) = -\frac{\rho\nu'}{6r},
\end{eqnarray}
\begin{eqnarray}
{\cal P}(r) = -\frac{\rho}{6r^2}\frac{\partial}{\partial
r}\left(r^2\frac{\partial \lambda}{\partial r}\right),
\end{eqnarray}
\begin{eqnarray}
\rho_{curv}(r) &=&
-\mu^2\left(1+\frac{2\rho}{\lambda_b}\right)\left(R_0''+\frac{2}{r}R_0'\right)\frac{d^2{\cal
L}}{dR^2}-\varepsilon
g'(R)\mu^2\left[1+\frac{\mu^2}{2\lambda_b}\left(R_0''+\frac{2}{r}R_0'\right)\frac{d^2{\cal
L}}{dR^2}-\frac{\rho}{2\lambda_b}\left(1-\frac{12\nu'}{r}\right)\right]\nonumber\\
&-&\varepsilon
g'(R)\mu^2\left(1+\frac{\mu^2}{2\lambda_b}+\frac{2\rho}{\lambda_b}\right)\frac{1}{r^2}\frac{\partial}{\partial
r}\left(r^2\frac{\partial \lambda}{\partial r}\right),
\end{eqnarray}
and
\begin{eqnarray}
\Lambda_4 =
\frac{\kappa_{5}^{2}}{2}\left[\Lambda^{(5)}+\frac{1}{6}\kappa_{5}^{2}\lambda_b^2\right],\label{A3}
\end{eqnarray}
which in the limit of ${\cal L}(R) = R$, we have $\rho_{curv}(r) =
0$. Multiplying equation (\ref{b14}) by $r^2$ and integrating from
0 to $r$ yields
\begin{eqnarray}
\left(1+\frac{\lambda_b}{6}\kappa_{5}^4\mu^2\right)\frac{1}{2}\left(r^2\frac{\partial
\lambda}{\partial
r}\right)-\frac{\kappa_4^2}{8\pi}M(r)+\frac{1}{3}\Lambda_4r^3-\frac{\kappa_4^2}{8\pi}M_{curv}(r)
-\frac{\kappa_4^2}{8\pi}{
M}_{DGP}(r)-\frac{\kappa_4^2}{8\pi}{M}_{RS}(r) = 0,\label{b15}
\end{eqnarray}
which the total baryonic mass and the geometrical masses of the
system are given by
\begin{eqnarray}
M(r) = 4\pi \int^r_0\rho(r') r^{'2}dr^{'},\label{b16}
\end{eqnarray}
\begin{eqnarray}
M_{curv}(r) = 4\pi \int^r_0\rho_{curv}(r')
r^{'2}dr^{'},\label{b17}
\end{eqnarray}
and
\begin{eqnarray}
\kappa_4^2{M}_{DGP}(r) = 8\pi\kappa_5^4\mu^2 \int^r_0 [{\cal
P}(r')+3{\cal U}(r')] r^{'2}dr^{'},\label{b19}
\end{eqnarray}
\begin{eqnarray}
\kappa_4^2{M}_{RS}(r) = \frac{48\pi}{\kappa_4^2\lambda_b} \int^r_0
{U}(r') r^{'2}dr^{'}.\label{b18}
\end{eqnarray}
Multiplying equation (\ref{b15}) by $\frac{dM(r)}{r}$ and
integrating from 0 to $R$, we finally obtain the generalized
virial theorem in modified DGP scenario as
\begin{eqnarray}
\left(1+\frac{\lambda_b}{6}\kappa_{5}^4\mu^2\right)2K+W+\frac{1}{3}\Lambda_4I+W_{curv}+W_{DGP}+W_{RS}
= 0,\label{Virial}
\end{eqnarray}
where
\begin{eqnarray}
W = -\frac{\kappa_4^2}{8\pi}\int^R_0\frac{M(r)}{r}
dM(r),\label{b20}
\end{eqnarray}
and
\begin{eqnarray}
I = \int^R_0r^2dM(r),\label{b21}
\end{eqnarray}
are the gravitational potential energy and the moment of inertia
of the system, respectively. We can also define the gravitational
potential energy due to the geometrical masses as
\begin{eqnarray}
{W}_{curv} = -\frac{\kappa_4^2}{2}\int^R_0{M}_{curv}(r)\rho r
dr,\label{b24}
\end{eqnarray}
\begin{eqnarray}
{W}_{DGP} = -\frac{\kappa_4^2}{2}\int^R_0{M}_{DGP}(r)\rho r
dr,\label{b23}
\end{eqnarray}
\begin{eqnarray}
{W}_{RS} = -\frac{\kappa_4^2}{2}\int^R_0{M}_{RS}(r)\rho r
dr.\label{b22}
\end{eqnarray}
By choosing ${\cal L}(R) = R$, we can also obtain the virial
theorem in warped DGP brane-world as \cite{Heydari}
\begin{eqnarray}
\left(1+\frac{\lambda_b}{6}\kappa_{5}^4\mu^2\right)2K+W+\frac{1}{3}\Lambda_4I+
W_{DGP}+W_{RS} = 0.\label{25}
\end{eqnarray}
At this point it is worth nothing that for $\mu = 0$, we have
$M_{DGP} = 0$ and $M_{curv} = 0$ so that the virial theorem in the
modified DGP brane-world is reduced to the virial theorem in the
RS brane scenario and the gravitational energy modified by
$W_{RS}$ which has its origin in the global bulk effect due to the
${\cal E}_{\mu\nu}$ term as \cite{Harko}
\begin{eqnarray}
2K+W+\frac{1}{3}\Lambda_4 I+W_{RS} = 0.\label{b26}
\end{eqnarray}
In the case $\lambda_b = \Lambda^{(5)} = 0$, the virial theorem in
modified DGP brane-world is reduced to the virial theorem in
DGP-inspired ${\cal L}(R)$ gravity scenario
\begin{eqnarray}
2K+W+W_{curv}+W_{DGP} = 0,\label{b27}
\end{eqnarray}
for ${\cal L}(R) = R$, we can also obtain the virial theorem in
DGP brane-world as \cite{Shahidi}
\begin{eqnarray}
2K+W+W_{DGP} = 0.\label{b28}
\end{eqnarray}
We note that in equations (\ref{b27}) and (\ref{b28}) $W_{RS} =
0$, since in a DGP brane-world with a Minkowski bulk space ${\cal
E}_{\mu\nu} = 0$. There is difference between our model with
references \cite{Harko} and \cite{Shahidi}. Here, the virial
theorem modified by $W_{RS}$, $W_{DGP}$ and $W_{curv}$ which the
first is due to the global bulk effect whereas the second term has
its origins in the induced gravity on the brane due to quantum
correction and the third term arises from higher-order corrections
on the scalar curvature over the brane. Also, in contrast to the
warped DGP model when the five-dimensional contribution is frozen
in the modified DGP model, the geometrical terms from the
four-dimensional ${\cal L}(R)$ gravity can be used to explain the
virial theorem mass discrepancy in clusters of galaxies. (see
Appendix B)

Now, we define the following geometrical mass
\begin{eqnarray}
{\cal M}(r) = M_{RS}(r)+M_{DGP}(r)+M_{curv}(r),\label{Mass}
\end{eqnarray}
so that
\begin{eqnarray}
{\cal W} = -\frac{\kappa_4^2}{2}\int^R_0{\cal M}(r)\rho r
dr.\label{}
\end{eqnarray}
In order to obtain a relation between the virial mass and the
geometrical mass ${\cal M}(r)$, we introduce the radii $R_{V}$ ,
$R_I$ and ${\cal R}$ as \cite{Jackson}
\begin{eqnarray}
R_V = \frac{M^2(r)}{\int_0^R \frac{M(r)}{r}dM(r)},\label{b288}
\end{eqnarray}
\begin{eqnarray}
R_I^2 = \frac{\int_0^R r^2 dM(r)}{M(r)},\label{b29}
\end{eqnarray}
\begin{eqnarray}
{\cal R} = -\frac{\kappa_4^2}{8\pi}\frac{{\cal M}^2(r)} {{\cal
W}},\label{b30}
\end{eqnarray}
where $R_{V}$ is the virial radius and ${\cal R}$ is defined as
the geometrical radius of the clusters of galaxies. By defining
the virial mass as
\begin{eqnarray}
2K = -\frac{\kappa_4^2}{8\pi}\frac{M_V^2}{R_V},\label{b31}
\end{eqnarray}
and using the relations
\begin{eqnarray}
W = -\frac{\kappa_4^2}{8\pi}\frac{M^2}{R_V},\hspace{.5 cm}I =
MR_I^2,\label{b32}
\end{eqnarray}
the generalized virial theorem (\ref{Virial}) is simplified as
\begin{eqnarray}
\left(1+\frac{\lambda_b}{6}\kappa_{5}^4\mu^2\right)\left(\frac{M_V}{M}\right)^2
= 1-\frac{8\pi\Lambda_4}{3\kappa_4^2}\frac{
R_VR_I^2}{M}+\left(\frac{\cal M}{M}\right)^2\left(\frac{R_V}{\cal
R}\right).\label{b33}
\end{eqnarray}
Since the dark matter provides the main mass contribution to
clusters, one can ignore the mass contribution of the baryonic
mass in the clusters and estimate the total mass of the cluster by
${\cal M}$. As a result, we conclude that
\begin{eqnarray}
M_V(r) \simeq {\cal M}(r)\left(\frac{R_{V}}{{\cal
R}}\right)^{1/2}.\label{b34}
\end{eqnarray}
This shows that the virial mass is proportional to the geometrical
mass.

\section{Estimating the geometrical mass ${\cal M}(r)$}

In clusters of galaxies the most of the baryonic mass is in the
form of the intra-cluster gas. The following equation provides a
reasonably good description of the observational date \cite{Y}
\begin{equation}\label{c1}
\rho_g(r) =
\rho_0\left(1+\frac{r^2}{r_c^2}\right)^{\frac{-3\beta}{2}},
\end{equation}
where $r_c$ is the core radius, and $\rho_0$ and $\beta$ are
cluster independent constants.

The observed x-ray emission from the hot ionized intra-cluster gas
is usually interpreted by assuming that the gas is in isothermal
equilibrium. Therefore we assume that the gas is in equilibrium
state having the equation of state $p_g(r) = \frac{k_B T_g}{\mu
m_p}\rho_g(r)$, where $k_B$ is Boltzmann constant, $T_g$ is the
gas temperature, $\mu = 0.61$ is the mean atomic weight of the
particles in the gas cluster and $m_p$ is the proton mass. Thus,
with the use of the Jeans equation \cite{book}, the total mass
distribution can be obtained as \cite{Harko, Y}
\begin{equation}\label{c2}
M_{tot}(r) = -\frac{8\pi k_B T_g}{\mu m_p \kappa_4^2} r^2
\frac{d}{dr}\ln\rho_g(r).
\end{equation}
Now, substitution of the mass density of the cluster gas in
equation (\ref{c1}) gives the total mass profile inside the
cluster \cite{Harko, Y}
\begin{equation}\label{c3}
M_{tot}(r) = \frac{24\pi k_BT_g\beta} {\mu m_p\kappa_4^2 }
\frac{r^3} {r^2 + r_c^2}.
\end{equation}
On the other hand, using equation (\ref{b16})-(\ref{b19}) we can
obtain another expression for the total mass
\begin{equation}\label{c4}
\frac{dM_{tot}(r)}{dr} = 4\pi r^2\rho_g(r) + 4\pi
r^2\rho_{curv}(r)+\frac{8\pi\kappa_5^4\mu^2}{\kappa_4^2}\left[{\cal
P}(r)+3{\cal U}(r)\right]r^2+
\frac{48\pi}{\kappa_4^2\lambda_b}U(r)r^2.
\end{equation}
Since we have estimated the quantities $M_{tot}(r)$ and
$\rho_g(r)$, the expression for the geometric mass density can be
readily obtained
\begin{equation}\label{c5}
\frac{6k_BT_g\beta} {\mu m_p \kappa_4^2} \frac{r^2+3r_c^2}{
(r^2+r_c^2)^2} =
\rho_0\left(1+\frac{r^2}{r_c^2}\right)^{-3\beta/2}+\rho_{curv}(r)+\frac{2\kappa_5^2\mu^2}{\kappa_4^4}\left[{\cal
P}(r)+3{\cal U}(r)\right]+\frac{12}{\kappa_4^2\lambda_b}U(r).
\end{equation}
Finally, substituting above equation into equation (\ref{Mass}),
in the limit $r>>r_c$ considered here, we obtain the following
geometrical mass
\begin{equation}\label{c6}
{\cal M}(r) \simeq \left[\frac{24\pi k_BT_g\beta} {\mu m_p
\kappa_4^2}-4\pi\rho_0r_c^{3\beta}\frac{r^{2-3\beta}}{3(1-\beta)}\right]
r.
\end{equation}
Observations show that the intra cluster gas has a small
contribution to the total mass \cite{W, Z, X}, thus we can neglect
the contribution of the gas to the geometrical mass and rewrite
equations (\ref{c6}) as
\begin{equation}\label{c7}
{\cal M}(r) \simeq \left(\frac{24\pi k_BT_g\beta} {\mu m_p
\kappa_4^2}\right) r.
\end{equation}
To estimate the value of ${\cal M}(r)$, we first note that $k_B
T_g \approx 5 KeV$ for most clusters. The virial radius of the
clusters of galaxies is usually assumed to be $r_{200}$,
indicating the radius for which the energy density of the cluster
becomes $\rho_{200} = 200 \rho_{cr}$, where $\rho_{cr} =
4.6975\times10^{-30}h^2_{50} gr/cm^{3}$ \cite{Y}. Using equation
(\ref{c6}) we find
\begin{equation}\label{c8}
r_{cr} = 91.33 {\beta}^{1/2}\left(\frac{k_BT_g}{5
KeV}\right)^{\frac{1}{2}} h^{-1}_{50} Mpc.
\end{equation}
The total geometrical mass corresponding to this value is
\begin{equation}\label{c9}
{\cal M}{(r)} =
4.83\times10^{16}{\beta}^{3/2}\left(\frac{k_BT_g}{5
KeV}\right)^{\frac{1}{2}} h^{-1}_{50} M_{\odot},
\end{equation}
which is consistent with the observational values for the virial
mass of clusters \cite{Y}.

\section{Radial velocity dispersion in galactic clusters}

The virial mass can be expressed in terms of the characteristic
velocity dispersion $\sigma_1$ as \cite{Z}
\begin{equation}\label{d0}
M_V = \frac{3}{G}\sigma_1^2 R_V.
\end{equation}
By assuming that the velocity distribution in the cluster is
isotropic, we have $\langle{v^2}\rangle = \langle{v_r^2}\rangle +
\langle{v_\theta^2}\rangle + \langle{v_\varphi^2}\rangle =
3\langle{v_r^2}\rangle = 3\sigma_r^2$, with $\sigma_1$ and
$\sigma_r$ are related by $3\sigma^2_1 = \sigma^2_r$.  In order to
derive the radial velocity dispersion $\sigma_r^2$ for clusters of
galaxies in modified DGP model we start from equation (\ref{b6})
as
\begin{equation}\label{d1}
\frac{d} {dr} (\rho\sigma_r^2 ) + \frac{1} {2} \rho\lambda' = 0.
\end{equation}
On the other hand, by neglecting the cosmological constant the
Einstein field equation (\ref{b14}) becomes
\begin{eqnarray}
\left(1+\frac{\lambda_b}{6}\kappa_{5}^4\mu^2\right)\frac{1}{2r^2}\frac{\partial}{\partial
r}\left(r^2\frac{\partial \lambda}{\partial r}\right) =
\frac{\kappa_4^2}{2}\rho(r)+\frac{\kappa_4^2}{2}\rho_{curv}(r)+{\kappa_5^4\mu^2}\left[{\cal
P}(r)+3{\cal
U}(r)\right]+\frac{6}{\kappa_4^2\lambda_b}U(r).\label{d3}
\end{eqnarray}
Integrating, we obtain
\begin{eqnarray}
\left(1+\frac{\lambda_b}{6}\kappa_{5}^4\mu^2\right)\frac{1}{2}\left(r^2\frac{\partial
\lambda}{\partial r}\right) = \frac{\kappa_4^2}{8\pi
}M(r)+\frac{\kappa_4^2}{8\pi }{\cal M}(r)+{{\cal C}_1},\label{d4}
\end{eqnarray}
where ${\cal C}_1$ is a constant of integration.

By eliminating $\lambda'$ from equations (\ref{d1}) and
(\ref{d4}), it follows that the radial velocity dispersion of the
galactic cluster satisfies the following differential equation
\begin{equation}\label{d5}
\left(1+\frac{\lambda_b}{6}\kappa_{5}^4\mu^2\right)\frac{d}{dr}(\rho\sigma_r^2)
= -\frac{\kappa_4^2M(r)}{8\pi r^2}\rho(r)-\frac{\kappa_4^2{\cal
M}(r)}{8\pi r^2}\rho(r)-\frac{{\cal C}_1}{r^2}\rho(r),
\end{equation}
with the general solution given by
\begin{eqnarray}\label{d6}
(1+\frac{\lambda_b}{6}\kappa_{5}^4\mu^2)\sigma_r^2 =
-\frac{1}{\rho}\int\frac{\kappa_4^2M(r)}{8\pi r^2}\rho(r)dr
-\frac{1}{\rho}\int\frac{\kappa_4^2{\cal M}(r)}{8\pi
r^2}\rho(r)dr-\frac{1}{\rho}\int\frac{{\cal
C}_1}{r^2}\rho(r)dr-\frac{{\cal C}_2}{\rho},
\end{eqnarray}
where ${\cal C}_2$ is a constant of integration.

For most clusters $\beta\geq\frac{2}{3}$ and therefore, in the
limit $r>> r_c$, the gas density profile (\ref{c1}) is given by
$\rho_g(r) = \rho_0 \left(\frac{r}{r_c}\right)^{-3\beta} $. Now,
substituting this relation for the gas density profile and
equation (\ref{c7}) into equation (\ref{d6}), for $\beta\neq 1$ we
find
\begin{equation}\label{d7}
\sigma_r^2 =
\frac{1}{\left(1+\frac{\lambda_b}{6}\kappa_{5}^4\mu^2\right)}\left[
-\frac{\rho_0\kappa_4^2}{12(1-\beta)(1-3\beta)}r^2\left(\frac{r}{r_c}\right)^{-3\beta}+\frac{k_BT_g}{\mu
m_p}+\frac{{\cal C}_1}{(1+3\beta)}\frac{1}{r}-\frac{{\cal
C}_2}{\rho_0}\left(\frac{r}{r_c}\right)^{3\beta}\right],
\end{equation}
and for $\beta = 1$ we have
\begin{equation}\label{d8}
\sigma_r^2 =
\frac{1}{\left(1+\frac{\lambda_b}{6}\kappa_{5}^4\mu^2\right)}\left[\frac{\rho_0
\kappa_4^2 }{8}\frac{\ln r}{r}+\frac{k_BT_g}{\mu m_p}+\frac{{\cal
C}_3}{r}+\frac{{\cal
C}_2}{\rho_0}\left(\frac{r}{r_c}\right)^{3}\right],
\end{equation}
with $4{\cal C}_3 = {\cal C}_1+\frac{\rho_0 \kappa_4^2 }{8}$. As
we noted before for $\Lambda^{(5)}=\lambda_b=0$, the modified DGP
model reduces to the DGP model and equation (\ref{d8}) for $\beta
= 1$ reduces to \cite{Shahidi}
\begin{equation}\label{d9}
\sigma_r^2 = \frac{\rho_0 \kappa_4^2 }{8}\frac{\ln
r}{r}+\frac{k_BT_g}{\mu m_p}+\frac{{\cal C}_3}{r}+\frac{{\cal
C}_2}{\rho_0}\left(\frac{r}{r_c}\right)^{3}.
\end{equation}
In figure 3 we have plotted the radial velocity dispersion for the
cluster NGC 5813. The numerical values of it are in the ranges
$\beta = 0.766$, $r_c = 25 Kpc$, $k_BT_g = 0.52 Kev$, $r_{200} =
0.87 Mpc$ \cite{Y} and the radial velocity is about $240 km/s$
\cite{An}. As one can see the radial velocity dispersion in
modified DGP brane-world is compatible with the observed profiles
and is the same with the warped DGP brane-world model.

\begin{figure}
\centerline{\begin{tabular}{ccc}
\epsfig{figure=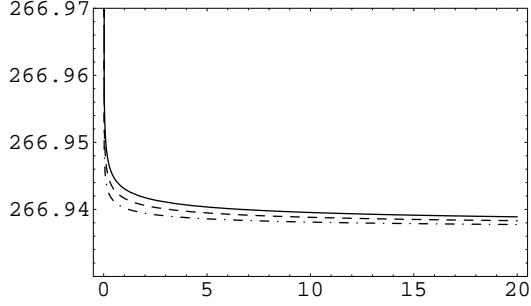,width=7cm}
\end{tabular} } \caption{\footnotesize  The radial velocity
dispersion in warped DGP brane-${\cal L}(R)$ model for the NGC
5813 cluster with ${\cal C}_1 = 0.503, 1.005\times 10^{8},
2.011\times 10^{8}\hspace{0.1cm}M_{\odot}$ and ${\cal C}_2 = 0.02,
0.03, 0.04\hspace{0.1cm}{M_{\odot}^2}/{Kpc}^{4}$ for solid, dashed
and dot-dashed curves, respectively.}\label{fig1}
\end{figure}

\section{Conclusions}

In this paper we have obtained a comprehensive version of the
virial theorem within the context of the modified DGP brane-world
model which can be reduced to the virial theorem in RS, DGP,
warped DGP brane-world models and ${\cal L}(R)$ gravity theories.
To derive the generalized virial theorem, we have used a method
based on the collision-less Boltzmann equation. For this end we
have also considered an isolated, spherically symmetric cluster,
situated in a space with metric given by equation (\ref{24}) and
described the galaxies, which are treated as identical,
collision-less point particles, by a distribution function which
obeys the general relativistic Boltzmann equation. It has been
shown that the geometrical mass which is due to the non-local bulk
effect and the induced gravity from the higher-order corrections
of the scalar curvature on the brane provides an effective
contribution to the gravitational energy, equation (\ref{b15}),
which may be used to explain the well-known virial theorem mass
discrepancy in clusters of galaxies. The virial theorem allows to
remove the virial mass discrepancy, by showing that the virial
mass $M_V$ is proportional to the geometrical mass ${\cal M}$,
${M}_{V} \simeq {\cal M}(\frac{R_V}{\cal R})^{\frac{1}{2}}$. The
geometrical mass can be directly related to the observed virial
mass, and for typical clusters of galaxies, it is of the same
order of magnitude as the virial mass. This shows that at the
extra galactic scale the geometrical mass plays the role of what
is conventionally called the dark matter.

We have also obtained the radial velocity dispersion profile of
clusters and used the observed radial velocity dispersion for the
cluster NGC 5813 as an example to show that our model can account
for the velocity dispersion of clusters. Finally, it is worth
noting that the advantage of the modified DGP model is when the
five-dimensional contribution is frozen. In this case the
geometrical terms from the four-dimensional ${\cal L}(R)$ gravity
can be also used to explain the virial theorem mass discrepancy in
clusters of galaxies \cite{Boehmer}.

\section*{Acknowledgment}

We would like to thank the anonymous referee for invaluable
comments and criticisms.

\section*{Appendix A: Field equations in four-dimensional ${\cal L}(R)$ gravity}

In this appendix we derive the four-dimensional limit of equation
(\ref{15}). Substituting $\Lambda(R)$, $\Sigma(R)$, $\Theta(R)$
and $\Lambda_4(R)$ into equation (\ref{15}), we obtain
\begin{eqnarray}
&&\left[1+\frac{1}{6}\kappa_{5}^{4}\lambda_b\mu^2{\frac{d{\cal
L}}{dR}}\left(1+\frac{\mu^2}{2\lambda_b}(R\frac{d{\cal
L}}{dR}-{\cal L})\right)\right]G_{\mu\nu} =
\frac{1}{6}\kappa_{5}^{4}\lambda_b\left[1+\frac{\mu^2}{2\lambda_b}(R\frac{d{\cal
L}}{dR}-{\cal L})\right]T_{\mu\nu}-\frac{1}{2}\kappa_{5}^{2}\Lambda^{(5)}g_{\mu\nu}\nonumber\\
&-&\frac{1}{12}\kappa_{5}^{4}\lambda_b^2g_{\mu\nu}\left[1+\frac{\mu^2}{\lambda_b}(R\frac{d{\cal
L}}{dR}-{\cal L})+\frac{\mu^4}{4\lambda_b^2}(R\frac{d{\cal
L}}{dR}-{\cal L})^2\right]+
\frac{1}{6}\kappa_{5}^{4}\lambda_b\mu^2\left[1+\frac{\mu^2}{2\lambda_b}(R\frac{d{\cal
L}}{dR}-{\cal L})\right] \nonumber\\&\times& \left[\nabla_\mu
\nabla_\nu{\left(\frac{d{\cal
L}}{dR}\right)}-g_{\mu\nu}\nabla_\beta
\nabla^\beta\left(\frac{d{\cal L}}{dR}\right)\right]
+\kappa_{5}^{4}\pi_{\mu\nu}^{(T)}+\kappa_{5}^4\pi_{\mu\nu}^{(\Theta)}
+\kappa_{5}^{4}\left(\mu^2\frac{d{\cal
L}}{dR}\right)^2\pi_{\mu\nu}^{(G)}\nonumber\\
&-&\kappa_{5}^{4}\left(\mu^2\frac{d{\cal L}}{dR}\right){
K}^{(T)}_{\mu\nu\rho\sigma}G^{\rho\sigma}
-\kappa_{5}^{4}\left[\mu^2\frac{d{\cal L}}{dR}
G^{\rho\sigma}-T^{\rho\sigma}\right]{
K}^{(\Theta)}_{\mu\nu\rho\sigma}-{\cal E}_{\mu\nu}.\label{A1}
\end{eqnarray}
Now we can recover the standard four-dimensional ${\cal L}(R)$
gravity from above equation in the limit $\kappa_5\rightarrow0$,
while keeping the Newtonian gravitational constant $\kappa_4^2 =
\frac{1}{6}\kappa_{5}^{4}\lambda_b$ finite \cite{maeda}
\begin{eqnarray}
&&\frac{1}{6}\kappa_{5}^{4}\lambda_b\mu^2\frac{d{\cal
L}}{dR}G_{\mu\nu} =
\frac{1}{6}\kappa_{5}^{4}\lambda_bT_{\mu\nu}-\Lambda_4g_{\mu\nu}-\frac{1}{12}\kappa_{5}^{4}\lambda_b\mu^2\left(R{\frac{d{\cal
L}}{dR}}-{\cal L}\right)g_{\mu\nu}\nonumber\\
&+&\frac{1}{6}\kappa_{5}^{4}\lambda_b\mu^2\left[\nabla_\mu
\nabla_\nu{\left(\frac{d{\cal
L}}{dR}\right)}-g_{\mu\nu}\nabla_\beta
\nabla^\beta\left(\frac{d{\cal L}}{dR}\right)\right],\label{A2}
\end{eqnarray}
where
\begin{eqnarray}
\Lambda_4 =
\frac{\kappa_{5}^{2}}{2}\left[\Lambda^{(5)}+\frac{1}{6}\kappa_{5}^{2}\lambda_b^2\right],\label{A3}
\end{eqnarray}
neglecting the effective cosmological constant we can also rewrite
equation (\ref{A2}) as
\begin{eqnarray}
G_{\mu\nu} = \frac{1}{(\frac{d{\cal L})}{dR}}\left[
\kappa_4^2T_{\mu\nu}+T_{\mu\nu}^{(curv)}\right],\label{A4}
\end{eqnarray}
where $\mu^2 \equiv\frac{1}{\kappa_4^2}$ and the curvature fluid
energy-momentum tensor is defined as
\begin{eqnarray}
T_{\mu\nu}^{(curv)} = \frac{1}{2}\left({\cal L}-R{\frac{d{\cal
L}}{dR}}\right)g_{\mu\nu}+\left[\nabla_\mu
\nabla_\nu{\left(\frac{d{\cal
L}}{dR}\right)}-g_{\mu\nu}\nabla_\beta
\nabla^\beta\left(\frac{d{\cal L}}{dR}\right)\right],\label{A5}
\end{eqnarray}
which is exactly the four-dimensional field equation in ${\cal
L}(R)$ gravity theories \cite{L(R)}.

\section*{Appendix B: Virial theorem in ${\cal L}(R)$ gravity}

Our aim in this appendix is to obtain the four-dimensional limit
of the virial theorem in modified DGP model and to compare its
results with pure ${\cal L}(R)$ gravity theories. From equation
(\ref{b13}) the virial theorem in modified DGP model is given by
\begin{eqnarray}
&&\left(1+\frac{\lambda_b}{6}\kappa_{5}^4\mu^2\right)\frac{1}{2r^2}\frac{\partial}{\partial
r}\left(r^2\frac{\partial \lambda}{\partial r}\right) =
\frac{1}{12}\kappa_{5}^{4}\lambda_b\rho(r)-\Lambda_4+\frac{1}{6}\kappa_{5}^{4}\mu^2\rho\left(R_0''+\frac{2}{r}R_0'\right)\frac{d^2{\cal
L}}{dR^2}\nonumber\\
&-&\varepsilon g'(R)
\frac{1}{12}\kappa_{5}^{4}\lambda_b\mu^2\left[1-\frac{\rho}{2\lambda_b}+\frac{\rho}{\lambda_b}\frac{3\nu'}{r}\right]
-\varepsilon
g'(R)\frac{1}{6}\kappa_{5}^{4}\lambda_b\mu^2\left(1+\frac{\mu^2}{2\lambda_b}+\frac{\rho}{\lambda_b}\right)\frac{1}{2r^2}\frac{\partial}{\partial
r}\left(r^2\frac{\partial \lambda}{\partial r}\right)\nonumber\\
&-&\frac{1}{6}\kappa_{5}^{4}\lambda_b\mu^2\left(1+\frac{\mu^2}{2\lambda_b}\varepsilon
g'(R)\right)\left(R_0''+\frac{2}{r}R_0'\right)\frac{d^2{\cal
L}}{dR^2}+{\kappa_5^4\mu^2}[{\cal P}(r)+3{\cal
U}(r)]+\frac{6}{\kappa_4^2\lambda_b}U(r),\label{B1}
\end{eqnarray}
in the limit $\kappa_5\rightarrow0$, with keeping the Newtonian
gravitational constant $\kappa_4^2 = \frac{1}{6}\kappa_5^4
\lambda_b$ finite, we obtain
\begin{eqnarray}
&&\frac{1}{6}\kappa_{5}^4\lambda_b\mu^2\frac{1}{2r^2}\frac{\partial}{\partial
r}\left(r^2\frac{\partial \lambda}{\partial r}\right) =
\frac{1}{12}\kappa_{5}^{4}\lambda_b\rho(r)-\Lambda_4-\varepsilon
g'(R) \frac{1}{12}\kappa_{5}^{4}\lambda_b\mu^2 -\varepsilon
g'(R)\frac{1}{6}\kappa_{5}^{4}\lambda_b\mu^2\frac{1}{2r^2}\frac{\partial}{\partial
r}\left(r^2\frac{\partial \lambda}{\partial r}\right)\nonumber\\
&-&\frac{1}{12}\kappa_{5}^{4}\lambda_b\mu^2\left(R_0''+\frac{2}{r}R_0'\right)\frac{d^2{\cal
L}}{dR^2},\label{B2}
\end{eqnarray}
neglecting the effective cosmological constant, $\Lambda_4$, we
can rewrite the above equation as
\begin{eqnarray}
\left(1+\varepsilon
g'(R)\right)\frac{1}{2r^2}\frac{\partial}{\partial
r}\left(r^2\frac{\partial \lambda}{\partial r}\right) =
\frac{\kappa_{4}^{2}}{2}\rho(r)-\frac{1}{2}\varepsilon
g'(R)-\frac{1}{2}\left(R_0''+\frac{2}{r}R_0'\right)\frac{d^2{\cal
L}}{dR^2},\label{B3}
\end{eqnarray}
where $\mu^2 \equiv\frac{1}{\kappa_4^2}$. As mentioned before we
present $\frac{d{\cal L}}{dR}$ as $\frac{d{\cal L}}{dR} =
1+\varepsilon g'(R)$ where $\varepsilon$ is a small quantity and
$g'(R)$ describes the modifications of the geometry due to the
presence of the tensor $\Theta_{\mu\nu}$. Therefore, we have
\begin{eqnarray}
\frac{1}{2r^2}\frac{\partial}{\partial r}\left(r^2\frac{\partial
\lambda}{\partial r}\right) \simeq 4\pi G\rho(r)+4\pi
G\rho_{curv}(r),\label{B4}
\end{eqnarray}
where
\begin{eqnarray}
4\pi G\rho_{curv}\simeq -4\pi G\rho(r)\varepsilon
g'(R)+\left[-\varepsilon
g'(R)-\frac{1}{2}\left(R_0''+\frac{2}{r}R_0'\right)\frac{d^2{\cal
L}}{dR^2}\right]\left(1+\varepsilon g'(R)\right)^{-1},\label{B5}
\end{eqnarray}
is the geometrical energy density. Equation {(\ref{B4})} is
exactly equation (17) in reference \cite{Boehmer}. Also the
resulting virial theorem from equation (\ref{B4}) is given by
\begin{eqnarray}
2K+W+W_{curv} = 0,\label{B6}
\end{eqnarray}
where
\begin{eqnarray}
{W} = -\frac{\kappa_4^2}{8\pi}\int^R_0\frac{M(r)}{r} d
M(r),\label{B7}
\end{eqnarray}
\begin{eqnarray} {W}_{curv} =
-\frac{\kappa_4^2}{2}\int^R_0{M}_{curv}(r)\rho r dr.\label{B8}
\end{eqnarray}

\end{document}